\documentclass[aps, prd, letterpaper, amsmath, amssymb, superscriptaddress, twocolumn, floatfix, nofootinbib,nobibnotes]{revtex4-1}
\usepackage[pdftex]{graphicx}
\usepackage[pdftex, pdfstartview={FitH}, pdfnewwindow=true, colorlinks=true, citecolor=blue, filecolor=blue, linkcolor=blue, urlcolor=blue, pdfpagemode=UseNone, bookmarks=false]{hyperref}
\usepackage{comment}

\newcommand{\Tr}{\ensuremath{\mbox{Tr}\,} }
\newcommand{\tr}{\ensuremath{\mbox{tr}\;}}
\newcommand{\beq}{\begin{equation}}
\newcommand{\eeq}{\end{equation}}
\newcommand{\psib}{\ensuremath{\overline{\psi}}}
\newcommand{\Psib}{\ensuremath{\overline{\Psi}}}

\newcommand{\cS}{\ensuremath{\mathcal S}}
\newcommand{\cG}{\ensuremath{\mathcal G}}
\newcommand{\cT}{\ensuremath{\mathcal T}}

\begin{document}
\title{Topology and strong four fermion interactions in four dimensions}

\author{Simon Catterall }
\email{smcatter@syr.edu}
\author{Nouman Butt}
\affiliation{Department of Physics, Syracuse University, Syracuse, New York 13244, United States}

\date{1 March 2018}

\begin{abstract}
We study massless fermions interacting through a particular four fermion term in four
dimensions. Exact symmetries prevent the generation of bilinear fermion mass terms. We determine the structure of the
low energy effective action for the auxiliary field needed to generate the four fermion term and
find it has an novel structure that admits topologically non-trivial defects with non-zero Hopf invariant. We show
that fermions propagating in such a background pick up a mass without breaking symmetries. Furthermore
pairs of such defects experience a logarithmic interaction.  We argue that a phase transition separates
a phase where these defects proliferate from a broken phase where they are bound tightly. We conjecture that
by tuning one additional operator the broken phase can be eliminated with a single BKT-like phase transition
separating the massless from massive phases.
\end{abstract}

\maketitle

\section{Introduction}
In this paper we construct a continuum theory of strongly interacting fermions in four dimensions 
in which exact symmetries prohibit the appearance of mass terms. We argue that the fermions nevertheless acquire masses at strong coupling by virtue of
their interactions with a non-trivial vacuum corresponding to a symmetric four fermion condensate. 
Our work points out the existence of new classes of theories of strongly interacting fermions
which may be important in the search for candidate theories of BSM physics.

Furthermore, we show that
the theory when discretized yields a staggered fermion lattice theory which has been the focus of several
recent studies both in the particle physics and condensed matter communities \cite{Ayyar:2014eua,Catterall:2015zua,
Ayyar:2015lrd, Ayyar:2016lxq,Catterall:2016dzf,He:2016sbs,Schaich:2017czc} in both three and four dimensions. The numerical work in three
dimensions is consistent with the absence of symmetry breaking bilinear condensates for all values of the four fermi coupling. The model nevertheless has a two phase structure with a continuous phase transition with non-Heisenberg
exponents separating a massless
phase from a phase with a symmetric four fermion condensate and massive fermions. Progress in
understanding the nature of this phase diagram was given recently in \cite{You:2017ltx}.
In four dimensions it appears that a very narrow symmetry broken phase emerges between the massless and massive
phases. 

The ingredients of the theory are somewhat unusual; the fermions 
appear as components of a (reduced) K\"{a}hler-Dirac field and as a  consequence the
theory is invariant only under a diagonal subgroup of the Lorentz and flavor symmetries together with an
additional $SO(4)$ symmetry.  It is this reduced symmetry, which is enforced by the structure of the four fermion term, that plays a key role in prohibiting conventional Dirac mass terms.

Our paper offers a way to understand the structure of the four dimensional models from a continuum perspective where we will see that
that topological features of the continuum theory can play an important role.

\section{Four fermion theory}
To start consider a theory comprising 4 flavors of free massless Dirac fermion with (Euclidean) action
\beq
S=\int d^4x\, \psib^a\gamma_\mu \partial_\mu\psi^a(x)\eeq
This is invariant under the global symmetry $SO_{\rm Lorentz}(4)\times SU_{\rm flavor}(4)$.
To build the model of interest let us focus on the diagonal subgroup of the Lorentz symmetry
and an $SO(4)$ subgroup of the original $SU(4)$ flavor symmetry which we call $\cT$.
\beq
\cT=SO^\prime(4)={\rm diag}\left[SO_{\rm Lorentz}(4)\times SO_{\rm flavor}(4)\right]\eeq
Under this symmetry we may rewrite the action as
\beq
S=\int d^4x\, {\rm Tr}\left(\Psib \gamma_\mu \partial_\mu\Psi\right)\eeq
where we now treat the fermions as $4\times 4$ matrices and the trace operation $\Tr$ occurring here and throughout the paper acts only on the matrix indices associated
with the $\cT$ symmetry.
Actually, since the theory is massless we can decompose these matrices into two independent
components using the twisted chiral projectors:
\beq
\Psi_\pm=\frac{1}{2}\left(\Psi\pm\gamma_5\Psi\gamma_5\right)\eeq
and the fermion action can be reduced to two Dirac flavors with action
\beq
S=\int d^4x\,{\rm Tr}\left(\Psib_+ \gamma_\mu \partial_\mu \Psi_-\right)\eeq
Notice that this projection only commutes 
with the $SO(4)$ subgroup of the original $SU(4)$ flavor symmetry. 
In the appendix we show that this reduction  is equivalent to imposing the reality condition $\Psib=\Psi$ with action
\beq
S=\int d^4x\,{\rm Tr}\left(\Psi\gamma_\mu \partial_\mu \Psi\right)\eeq
The equation of motion that follows from this action  can be interpreted as the (reduced) K\"{a}hler-Dirac equation if one expands the
fermion matrices on products of Dirac gamma matrices \cite{Banks:1982iq}. For the model we want to discuss we will consider four copies of this
system by taking these matrix fermions to additionally transform in the fundamental representation of
an independent $SO(4)$ symmetry $\cS$  i.e $\Psi^\alpha\to R^{\alpha\beta}\Psi^\beta$ with $R$ an element
of $SO(4)$.

Up to this point everything we have done 
merely corresponds to a change of variables that serves to highlight a particular subgroup of the global symmetries -
the diagonal subgroup of the Lorentz and flavor symmetries. The field content of the model still corresponds to
8 flavors of massless Dirac fermions.
However this situation changes when I add four fermion interactions of the following form
\beq
\delta S=\frac{G^2}{4}\int d^4x\, \epsilon_{\alpha\beta\gamma\delta}\Tr\left(\Psi^\alpha\Psi^\beta\right)\Tr\left(\Psi^\gamma\Psi^\delta\right)\label{act2}\eeq
This interaction locks the Lorentz and flavor symmetries together and
ensures that the global symmetries $\cG$ of the theory are 
\beq
{\cal G}=\cT\times \cS = SO^\prime(4)\times SO(4)\eeq
It is of crucial importance to notice that the resultant
theory does {\it not} admit any bilinear mass terms since $\Tr \Psi^\alpha\Psi^\alpha=0$ and
any terms of the form $\Tr \Psi^\alpha\Psi^\beta$ break the symmetry $\cS$.

\section{Aside: connection to (reduced) staggered fermions}
The motivation for this work derives in part from recent numerical investigations of lattice models involving
four reduced staggered fermions interacting through the corresponding unique four fermion interaction. In this section
we will show that the continuum model described earlier when discretized naturally leads to those lattice models.
One way to discretize 
the continuum theory is to expand the fermion matrices on position dependent products of
Dirac gamma matrices \cite{Bock:1992yr}. Consider the original $\Psi$ 
\beq
\Psi(x)=\sum_{b}\gamma^{x+b}\chi(x+b)\eeq
where the components of the vector $b^i=0,1$ label points in the unit hypercube attached to site x in a 
four dimensional hypercubic lattice
and 
\beq
\gamma^{b}=\prod_i \left(\gamma_i\right)^{b_i}\eeq
Plugging this expansion into eqn.(6) and doing
the trace over the gamma matrices yields the free
reduced staggered fermion action comprising one single component lattice fermion at each lattice
site:
\beq
\sum_{x,\mu}\chi(x) \eta_\mu(x)\Delta_\mu \chi(x)\eeq
with $\Delta_\mu$ the symmetric difference operator and $\eta_\mu(x)=\left(-1\right)^{\sum_{i=0}^{\mu-1}x_i}$ the usual staggered fermion phase
\cite{Golterman:1984cy,vandenDoel:1983mf}.
Equipping each of these fields with an
index under the {\cal S} symmetry and adding the four fermion terms one arrives at
\beq
S_{\rm stag}=\sum_{x,\mu}\chi^a(x) \eta_\mu(x)\Delta_\mu \chi^a(x)+\frac{G^2}{4}\sum_x\epsilon_{abcd}
\chi^a\chi^b\chi^c\chi^d\eeq
which is precisely the action studied in \cite{Ayyar:2016lxq,Catterall:2016dzf}.
Thus we expect that the continuum arguments described in this paper can be applied to understand the
numerical results reported for this staggered fermion system. 

\section{Auxiliary field action}
As usual our subsequent analysis requires
replacing the four fermion term given in eqn.(7) 
by a Yukawa coupling to an auxiliary scalar field
\beq
S_0=\int d^4x\left[iG\phi^{\alpha\beta}_+(x)
\Tr \left(\Psi^\alpha\Psi^\beta\right)+\frac{1}{4}\left(\phi_+^{\alpha\beta}\right)^2\right]\label{start}\eeq
The auxiliary field is a antisymmetric matrix and satisfies a self-dual condition $\phi_+={\cal P}^+\phi$ where the
projector ${\cal P}_+$ is defined as
\beq
{\cal P}^+_{\alpha\beta\gamma\delta}=\frac{1}{2}\left(\delta_{\alpha\gamma}\delta_{\beta\delta}+\frac{1}{2}\epsilon_{\alpha\beta\gamma\delta}\right)\eeq
Notice that the original four fermion interaction can be written  as 
\beq
\left[{\rm Tr}\,\Psi^\alpha\Psi^\beta\right]_+^2=\frac{1}{4G^2}\left(\phi_+^{\alpha\beta}\right)^2\label{fourfermion}\eeq
This structure ensures that  $\phi_+$ transforms in the adjoint representation
under a $SU_+(2)$ subgroup of the $\cS$ symmetry $SO(4)=SU_+(2)\times SU_-(2)$. It is a singlet under
both $SU_-(2)$ and the internal $\cT$ symmetries (see the appendix for
more details).  Furthermore, it is easy to 
see that the eigenvalues of the resultant fermion operator 
come in complex conjugate pairs. In addition each eigenvalue is doubly degenerate
since the fermion operator also commutes with $SU_-(2)$. These facts ensure that the Pfaffian that results
from integration over the fermions is in fact real, positive definite.

\section{Effective Action}
Returning to eqn.~\ref{start} we now integrate out the fermions using positivity of the Pfaffian and consider the form of the one loop
effective action.
\beq
S_{\rm eff}=-\frac{1}{4}{\rm Tr}\,\ln{\left(-\Box+G^2\mu^2+G\gamma_\mu\partial_\mu\phi_+\right)}\label{effact}\eeq
where $\phi_+^2=\mu^2\,I$ and we have absorbed the explicit factor of i into the auxiliary field to render
$\phi_+$ hermitan.
Let us first consider the Coleman-Weinberg effective potential obtained by assuming a constant  auxiliary field 
\beq
V_{\rm eff}(\mu)=-\frac{1}{4}\Tr\ln{\left( \frac{-\Box+G^2\mu^2}{-\Box}\right)}+\mu^2\eeq
where we have subtracted off the value of $V_{\rm eff}$ at $G=0$ and added in the classical action for $\phi_+$.
If we expand the remainder in powers of $G$ it should be clear that $V_{\rm eff}$ develops a minimum away from the
origin for sufficiently large $G>G_c$. Thus naively one expects the system to enter a symmetry broken state
for some value of the four fermi coupling. This is the usual NJL scenario and in this case will
correspond to a breaking 
pattern $SU_+(2)\to U(1)$ corresponding to a vacuum manifold with the topology of $S^2$. 

Of course to understand the dynamics of the theory in more detail we need to compute the leading
terms in the effective action for $\phi_+$ for {\it non constant} fields. Expanding the latter on a suitable $4\times 4$
basis $T$ (see the appendix for more details) we find
\beq
\phi_+(x)=\sum_{a=1}^3 \phi_+^a(x) T_a=\sum_{a=1}^3 n^a(x) \sigma^a\otimes I\eeq
In this basis the fermion operator has a trivial dependence on $SU_-(2)$ and we will suppress it
in our subsequent analysis. For $G>G_c$ the field $n^a(x)$ obeys the $O(3)$ constraint
$n^an^a=1$. The effective action governing the fluctuations in $n^a(x)$ is now given by a derivative expansion of
\beq
-\frac{1}{4}{\rm Tr}\,\ln{\left(I+m\frac{\gamma_\mu\partial_\mu n^a\sigma^a}{-\Box+m^2}\right)}\eeq
where $m=G\mu$.
At leading order one encounters an $O(3)$ symmetric term quadratic in
the derivatives of $n^a(x)$ (see the appendix)
\beq
a(G) \int d^4x\,\left(\partial_\mu n^a\right)^2
\label{effaction}\eeq
However at higher orders in $1/m$ one also encounters an additional quartic 
term which can play an important role in understanding the possible phases of the theory.
\beq
b(G)\int d^4 x\,\left(\epsilon^{abc}\partial_\mu n^a \partial_\nu n^b\right)^2\label{skyrme}\eeq
The combination of these two terms defines the Fadeev-Skyrme model which is known to possess topologically stable field configurations which we will argue can play a role in the current theory.

The analysis of the dynamics is facilitated by a further change
of variables in which the $O(3)$ vector $n^a$ is replaced by a $SU(2)$ matrix field which rotates
$n^a\sigma^a$ to a fixed matrix - say $\sigma_3$.
\beq
n^a(x)\sigma^a=U^\dagger(x)\sigma_3U(x)
\label{map}\eeq
This has the immediate advantage that the nonlinear constraint
$n^an^a=1$ is simply replaced by the unitarity property of $U=e^{i\theta^a\sigma^a}$ with the angular
variables $\theta$'s unconstrained. 
Of course this mapping cannot be the whole
story since the manifold of $SU(2)$ is $S^3$ not $S^2$ and indeed it is easy to see that $n^a$ is
is invariant under {\it local} left multiplication of $U(x)$  by an element of $U(1)$:
\beq
U(x)\to e^{i\sigma_3 \beta(x)}U(x)
\label{U1}\eeq
The action is also manifestly invariant under right multiplication by a {\it global} $SU(2)$ rotation $U\to UG$.
Thus the final effective action for $U$ should respect both this global $SU(2)$ symmetry and the local $U(1)$ gauge
symmetry. We can make the local invariance explicit if we replace ordinary derivatives by covariant
derivatives with the leading term now being \beq
S_{\rm eff}=a(G)\int d^4x\, \tr\left[ \left(D_\mu U\right)^\dagger \left(D_\mu U\right)\right]+\ldots\label{leading}
\eeq
where 
$D_\mu=\partial_\mu+iA_\mu\sigma_3$ and $A_\mu$ is an abelian gauge field needed to enforce the
$U(1)$ symmetry given in eqn.~\ref{U1}. This action is classically equivalent to the original one.
However in this case one would also expect to find a Maxwell term corresponding to this exact local $U(1)$ invariance
\beq
\delta S_{\rm eff}=b(G)\int d^4x\, F_{\mu\nu}F_{\mu\nu}\eeq
Indeed, classically, the field strength can be expressed in terms of $O(3)$ vector $n$ \cite{vanBaal:2001jm} as
\beq
F_{\mu\nu}=n.\left(\partial_\mu n\times \partial_\nu n\right)\eeq
and we see that the Maxwell term just represents the higher order term in eqn.~\ref{skyrme}.

In this picture a conventional broken phase for the sigma model eg $n^a=\delta^{a3}$ leads to $U=I$ up to gauge
transformations and corresponds to a Higgs phase with photon mass $\sqrt{a(G)}$. Close
to $G_c$ the photon mass is large and the gauge field decouples from long distance physics so that
this regime is governed by the usual $O(3)$ sigma model action.

\section{Topological defects}
While the uniform phase is always a possible vacuum solution additional possibilities arise at strong coupling
where the quartic term plays a role.
Let us search for non-trivial field configurations. To try to keep
the action finite forces us to look for solutions  where $D_\mu U\to 0$ 
as $r\to\infty$ and corresponding to vanishing photon mass. This implies
\beq
\partial_\mu U=-iA_\mu\sigma_3 U\eeq
or
\beq
A_\mu=\frac{i}{2}\tr\left(\partial_\mu U U^\dagger\sigma_3\right)\eeq
The long distance contribution to the action of such a configuration 
is then determined by the Maxwell term
\beq
b(G)\int d^4x\,\frac{1}{4} \left(\tr\partial_\mu U\partial_\nu U^\dagger\sigma_3\right)^2\label{max}\eeq
A topological defect must then correspond to a $U(x)$ configuration that maps non-trivially at
infinity into the $S^2$ target space. Such a mapping exists, is termed the Hopf map, and corresponds to $\Pi_3(S^2)=Z$.
If we parametrize a general $U$ matrix as
\beq
U=\left(\begin{array}{cc}
\alpha_1+i\alpha_2& -\alpha_3+i\alpha_4\\
\alpha_3+i\alpha_4& \alpha_1-i\alpha_2\end{array}\right)\eeq
with $\sum_i\alpha_i^2=1$
then the simplest topological defect corresponds to setting $\alpha_i=\frac{x_i}{r}$ where $x_i$ are
the four dimensional coordinates.  This parametrization 
yields a $S^3\to S^3$ map but this is reduced to the Hopf map when $U$ fields which are  gauge equivalent are identified. A similar topological defect solution was
constructed in a four dimensional Yang-Mills-Higgs system in  \cite{He:2014eoa}.
The $\alpha_i$ correspond to trigonometric functions of angles
in four dimensional polar coordinates and it can easily
seen that the action given in eqn.~\ref{max} corresponding to such a defect diverges logarithmically with system size\footnote{For a Hopf defect the gauge field corresponds to a large gauge transformation}.
Furthermore the topological charge of this object can be obtained from the theta term corresponding to the $U(1)$ field.
\beq
\frac{1}{32\pi^2}\int d^4x\, \epsilon_{\mu\nu\rho\lambda}\tr\left(\partial_\mu U\partial_\nu U^\dagger\sigma_3\right)\tr
\left(\partial_\rho U\partial_\lambda U^\dagger\sigma_3\right)\eeq
Unlike the action this term does not diverge logarithmically since it may be recast as a Chern-Simons term
which can be computed on the boundary sphere at infinity.

While such a background corresponds asymptotically to a point on the vacuum manifold it clearly does
not break the $\cS$ symmetry since $<\sum_x \phi_+(x)>=0$.
Of course the key question is whether such defects can play a role in determining the phase structure
of the model. At first glance they should not - the logarithmically divergent
action corresponding to such defects will ensure that a
single defect is completely suppressed in the infinite volume limit.
This situation is analogous to the behavior of vortices in the two dimensional XY model which also
possess a log divergent action.  In the latter case
a configuration of finite action
can be constructed consisting of a vortex and anti-vortex. The action for such a configuration
depends logarithmically on the separation of the two vortices  which hence bind tightly together at low temperatures.
However since the entropy associated with a vortex also increases logarithmically with system size
a BKT phase transition develops as the temperature is raised and vortices unbind and populate the
ground state. 

We propose that a similar phenomena may occur in this four dimensional model - that is the ground state for $G\sim G_c$ consists of tightly bound Hopf-antiHopf defects. In such a scenario
the disordering effects of the defects are suppressed and one expects a conventional symmetry
broken (Higgs) phase to appear as has been observed in the
numerical simulations \cite{Ayyar:2016lxq,Schaich:2017czc}. However as the coupling is increased still further
the defects may unbind via another transition to populate and disorder the ground state. This condensate
of Hopf defects with $<\phi_+^2>\ne 0$ would then correspond to the four fermion condensate in the original four fermi model consistent with
eqn.~\ref{fourfermion}. An estimate for the critical coupling can be arrived at by comparing the entropy associated to the
location of a single defect $S\sim \ln{V}$ with its action $E\sim b(G)\ln{V}$ yielding $b(G)^{\rm crit}\sim 1$.

It is interesting to compute the fermion propagator in the background of such a defect.
Consider the
$\cS$-symmetric correlator
\begin{eqnarray}
G_F(x,y)&=&\tr\left\langle\Psi(x)\Psi(y)\right\rangle\nonumber\\
&=&\tr\left[\frac{-\gamma_\mu\partial_\mu+m\,n^a \sigma^a}{\left(-\partial_\mu^2+m^2+mP\right)}\right]\end{eqnarray}
where 
\beq
P=\gamma_\mu\left(\partial_\mu U^\dagger(x)\sigma_3 U(x)+U^\dagger(x)\sigma_3\partial_\mu U(x)\right)\eeq
and the trace is to be carried out over the $\cS$-indices.
Using the fact that  the covariant derivative vanishes far from
the core of the defect allows us to show that $P=0$ and the propagator in that region simplifies to
\beq
G_F(x,y)=\frac{-2\gamma_\mu\partial_\mu}{-\Box+m^2}\eeq
Thus the fermion acquires a mass $m=\mu G$ in the background of such a defect. This
gives a concrete realization of the mechanism discussed in \cite{BenTov:2015gra} and is consistent with strong
coupling expansions for staggered fermions \cite{Catterall:2016dzf}.

\section{BKT transition}
We have argued that the model possesses a conventional broken phase (or Higgs phase)
which gives way to a symmetric phase at stronger
coupling due to unbinding of topological defects.  Since mechanisms for giving fermions a mass are
quite different in the two regimes one might expect a discontinuous phase transition separates the broken
phase and the defect phase.
To obtain a true BKT-like transition requires one to pass directly between the massless and massive
symmetric phases. To effect such a scenario one can generalize the original four fermion
model to a true Higgs-Yukawa model by the addition of a kinetic term for the auxiilary field $\phi_+$. One can then
imagine tuning the coupling of this kinetic operator so as to cancel out the effects of the leading gradient term eqn.~\ref{leading}. This sets the photon mass to zero and eliminates the Higgs phase of the model. We conjecture that
in this limit a true single BKT transition separates the massless and massive phases.

\section{Summary}
We have argued that a particular four dimensional continuum theory possesses an interesting phase structure as a function
of the coupling to a particular four fermion interaction. For sufficiently
weak four fermi coupling we expect the theory to describe massless non-interacting fermions. As the
coupling is increased the system should undergo a NJL-like phase transition to a phase in which
the $SO(4)$ symmetry is spontaneously broken via a bilinear fermion condensate. In the auxiliary field picture this
phase is characterized by tightly bound pairs of Hopf defects and a non-zero expectation value
for the scalar field. As the coupling is increased further we argue that
these defects may unbind at a transition to populate and disorder the vacuum restoring
the symmetry. In the background of such defects the
fermions acquire a mass without breaking symmetries. This phase is interpreted as a four fermion condensate
in the original fields. We also argue that by an additional tuning of the kinetic energy the broken phase can
be eliminated and a single BKT transition would separate the massless from massive phases. 

The continuum theory we describe possesses an unusual Lorentz symmetry which is locked via the four
fermion interaction with an internal flavor symmetry. At weak coupling we expect the four fermi term to be
irrelevant and the IR description of the theory will correspond to sixteen flavors of free Majorana fermion
with the symmetry enhancing to the usual Lorentz and flavor symmetries. Correspondingly
the beta function for the four fermi coupling has an IR attractive fixed point at $G=0$. The transition to a phase of
broken symmetry is likely of the NJL type and hence the corresponding (IR unstable) fixed point
would lie in the universality class of the usual Higgs-Yukawa theory. However if an additional continuous transition 
were to separate this phase from the four fermion condensate phase this would correspond to a new strongly
coupled IR fixed point. This would be a fascinating prospect.  The BKT limit would correspond to a situation where
the two fixed fixed points bounding the broken phase merge into a single continuous transition.

We have also argued 
that this continuum theory naturally discretizes to yield a theory of strongly interacting
reduced staggered fermions. This lattice model has received some recent attention and the numerical
phase diagram that has been uncovered matches quite
closely with the gross features described in this paper. Indeed, in the condensed matter literature there has recently been a great deal of 
interest in models which are able to gap fermions without breaking symmetries using carefully chosen quartic interactions \cite{Fidkowski:2009dba}. This work has even been used
to revive an old approach to lattice chiral gauge theories due to Eichten and Preskill \cite{Eichten:1985ft} in which mirror states of a definite chirality can be gapped out of
an underlying vector like lattice theory
using four fermion interactions \cite{You:2014vea}. It will be interesting to see whether the current model can be generalized to implement such constructions.
Independent of this potential connection,  the possibility of new phases and critical points in strongly interacting fermion systems in
four dimensions is very interesting in its own right and we hope the current work stimulates further work in this area. 

\acknowledgments This work is supported in part by the U.S.~Department of Energy, Office of Science, Office of High Energy Physics, under Award Number DE-SC0009998. SMC would like to acknowledge useful conversations
with Shailesh Chandrasekharan and Cenke Xu.

\section{Appendix}

\subsection*{Obtaining the twisted Majorana form}

Setting $\Psib_+=C^{-1}\Psi_+^T C$ where $C$ is the charge conjugation operator the action
can be rewritten
\beq
S=\int d^4x\,{\rm Tr}\,\left(C^{-1}\Psi^T_+ C\, \gamma_\mu\partial_\mu \Psi_-\right)\label{act1}\eeq
Taking the transpose of this equation yields
\beq
S=\int d^4x\,{\rm Tr}\,\left(C^{-1}\Psi_-^TC\,\gamma_\mu\partial_\mu \Psi_+\right)\eeq
Adding these two expressions the action can be expressed entirely in terms of the field $\Psi=\Psi_+ + \Psi_-$. 
\beq
S=\int d^4x\,{\rm Tr}\,\left(C^{-1}\Psi^TC\,\gamma_\mu\partial_\mu \Psi\right)\eeq
But $C^{-1}\Psi^TC=\Psi$ if one expresses the matrix $\Psi$ as a sum over the Clifford algebra formed from the product of Dirac
gamma matrices so that the action in (twisted) Majorana form is simply
\beq
S=\int d^4x\,{\rm Tr}\,\left(\Psi\,\gamma_\mu\partial_\mu\Psi\right)\eeq

\subsection*{Changing basis to $SU(2)\times SU(2)$}

We can verify the mapping into the $O(3)$ non-linear sigma model by starting from an explicit 
$4\times 4$ basis for the hermitian self-dual field $\phi_+=\sum_{a=1}^3 \phi_+^a T_a$
\beq
T_1=\left(\begin{array}{cc}0&-i\sigma_1\\i\sigma_1&0\end{array}\right)\; T_2=\left(\begin{array}{cc}0&i\sigma_3\\-i\sigma_3&0\end{array}\right)\;
T_3=\left(\begin{array}{cc}\sigma_2&0\\0&\sigma_2\end{array}\right)\nonumber\eeq
These matrices clearly obey an $SU(2)$ algebra which is part of the original $SO(4)$ $\cS$ algebra and the self-dual condition
is clearly equivalent to the statement that $\phi_+$ transforms in the adjoint representation of that $SU(2)$. The other independent $SU(2)$ contained
in $\cS$ is given the
generators 
\beq
U_1=\left(\begin{array}{cc}0&-\sigma_2\\-\sigma_2&0\end{array}\right)\;U_2=\left(\begin{array}{cc}0&i\sigma_1\\-i\sigma_1&0\end{array}\right)\;
U_3=\left(\begin{array}{cc}\sigma_2&0\\0&-\sigma_2\end{array}\right)\nonumber\eeq\\
Using the  similarity transformation $P$ given by
\beq
P=\frac{1}{\sqrt{2}}\left(\begin{array}{cccc}
1&0&0&-1\\
i&0&0&i\\
0&-1&-1&0\\
0&-i&i&0
\end{array}\right)
\eeq
one can verify that the generators $T$ and $U$ take the form
\beq
T^a=\sigma^a\otimes I\quad{\rm and}\quad U^a=I\otimes \sigma^a\eeq
This makes it clear that $T^a$ (and hence $\phi_+$) are singlets under $SU_-(2)$.

\subsection*{Large mass expansion}
Starting from the expression 
\beq
S_{\rm eff}=-\frac{1}{4}{\rm Tr}\,\ln{\left(-\Box+m^2+m\gamma_\mu\partial_\mu n^a\sigma^a\right)}\eeq
we first subtract the contribution at $m=0$ and write
\beq
S_{\rm eff}=-\frac{1}{4}{\rm Tr}\,\ln{\left[\left(\frac{-\Box+m^2}{-\Box}\right)\left(I+\frac{m\gamma_\mu\partial_\mu n^a\sigma^a}{-\Box+m^2}\right)\right]}\eeq
The first factor inside the logarithm yields the effective potential previously described. So we focus on the second factor. Clearly one can imagine
expanding this term in powers of $1/m$. To yield a non zero result one must arrange for a non zero trace over products of Dirac gamma matrices and
Pauli matrices. The leading term clearly arises at second order in $1/m$ and is
\beq
\int d^4x\, \frac{\Lambda^4}{m^2} \left(\partial_\mu n^a\right)^2\eeq
where $\Lambda$ is a UV cut-off. This term is quite generic and would arise independent of the structure of the Yukawa term. The structure
of the quartic term
depends crucially on the interplay of the $SU(2)$ and Dirac structures.
\beq
\int d^4x\, \frac{\Lambda^4}{m^4}\left(\partial_\mu n^a \partial_\nu n^b\right)^2\eeq
Since these operators contain the cutoff the coefficients must be renormalized to yield a finite effective action. We will not attempt that process here but merely
note that the coefficients of the effective action will have an explicit dependence on the mass $m$ and hence coupling $G$ and we write
them as $a(G)$ and $b(G)$.

\bibliography{kd4f}

\begin{thebibliography}{18}%
\makeatletter
\providecommand \@ifxundefined [1]{%
 \@ifx{#1\undefined}
}%
\providecommand \@ifnum [1]{%
 \ifnum #1\expandafter \@firstoftwo
 \else \expandafter \@secondoftwo
 \fi
}%
\providecommand \@ifx [1]{%
 \ifx #1\expandafter \@firstoftwo
 \else \expandafter \@secondoftwo
 \fi
}%
\providecommand \natexlab [1]{#1}%
\providecommand \enquote  [1]{``#1''}%
\providecommand \bibnamefont  [1]{#1}%
\providecommand \bibfnamefont [1]{#1}%
\providecommand \citenamefont [1]{#1}%
\providecommand \href@noop [0]{\@secondoftwo}%
\providecommand \href [0]{\begingroup \@sanitize@url \@href}%
\providecommand \@href[1]{\@@startlink{#1}\@@href}%
\providecommand \@@href[1]{\endgroup#1\@@endlink}%
\providecommand \@sanitize@url [0]{\catcode `\\12\catcode `\$12\catcode
  `\&12\catcode `\#12\catcode `\^12\catcode `\_12\catcode `\%12\relax}%
\providecommand \@@startlink[1]{}%
\providecommand \@@endlink[0]{}%
\providecommand \url  [0]{\begingroup\@sanitize@url \@url }%
\providecommand \@url [1]{\endgroup\@href {#1}{\urlprefix }}%
\providecommand \urlprefix  [0]{URL }%
\providecommand \Eprint [0]{\href }%
\providecommand \doibase [0]{http://dx.doi.org/}%
\providecommand \selectlanguage [0]{\@gobble}%
\providecommand \bibinfo  [0]{\@secondoftwo}%
\providecommand \bibfield  [0]{\@secondoftwo}%
\providecommand \translation [1]{[#1]}%
\providecommand \BibitemOpen [0]{}%
\providecommand \bibitemStop [0]{}%
\providecommand \bibitemNoStop [0]{.\EOS\space}%
\providecommand \EOS [0]{\spacefactor3000\relax}%
\providecommand \BibitemShut  [1]{\csname bibitem#1\endcsname}%
\let\auto@bib@innerbib\@empty
\bibitem [{\citenamefont {Ayyar}\ and\ \citenamefont
  {Chandrasekharan}(2015)}]{Ayyar:2014eua}%
  \BibitemOpen
  \bibfield  {author} {\bibinfo {author} {\bibfnamefont {V.}~\bibnamefont
  {Ayyar}}\ and\ \bibinfo {author} {\bibfnamefont {S.}~\bibnamefont
  {Chandrasekharan}},\ }\href {\doibase 10.1103/PhysRevD.91.065035} {\bibfield
  {journal} {\bibinfo  {journal} {Phys. Rev.}\ }\textbf {\bibinfo {volume}
  {D91}},\ \bibinfo {pages} {065035} (\bibinfo {year} {2015})},\ \Eprint
  {http://arxiv.org/abs/1410.6474} {arXiv:1410.6474 [hep-lat]} \BibitemShut
  {NoStop}%
\bibitem [{\citenamefont {Catterall}(2016)}]{Catterall:2015zua}%
  \BibitemOpen
  \bibfield  {author} {\bibinfo {author} {\bibfnamefont {S.}~\bibnamefont
  {Catterall}},\ }\href {\doibase 10.1007/JHEP01(2016)121} {\bibfield
  {journal} {\bibinfo  {journal} {JHEP}\ }\textbf {\bibinfo {volume} {01}},\
  \bibinfo {pages} {121} (\bibinfo {year} {2016})},\ \Eprint
  {http://arxiv.org/abs/1510.04153} {arXiv:1510.04153 [hep-lat]} \BibitemShut
  {NoStop}%
\bibitem [{\citenamefont {Ayyar}\ and\ \citenamefont
  {Chandrasekharan}(2016{\natexlab{a}})}]{Ayyar:2015lrd}%
  \BibitemOpen
  \bibfield  {author} {\bibinfo {author} {\bibfnamefont {V.}~\bibnamefont
  {Ayyar}}\ and\ \bibinfo {author} {\bibfnamefont {S.}~\bibnamefont
  {Chandrasekharan}},\ }\href {\doibase 10.1103/PhysRevD.93.081701} {\bibfield
  {journal} {\bibinfo  {journal} {Phys. Rev.}\ }\textbf {\bibinfo {volume}
  {D93}},\ \bibinfo {pages} {081701} (\bibinfo {year} {2016}{\natexlab{a}})},\
  \Eprint {http://arxiv.org/abs/1511.09071} {arXiv:1511.09071 [hep-lat]}
  \BibitemShut {NoStop}%
\bibitem [{\citenamefont {Ayyar}\ and\ \citenamefont
  {Chandrasekharan}(2016{\natexlab{b}})}]{Ayyar:2016lxq}%
  \BibitemOpen
  \bibfield  {author} {\bibinfo {author} {\bibfnamefont {V.}~\bibnamefont
  {Ayyar}}\ and\ \bibinfo {author} {\bibfnamefont {S.}~\bibnamefont
  {Chandrasekharan}},\ }\href {\doibase 10.1007/JHEP10(2016)058} {\bibfield
  {journal} {\bibinfo  {journal} {JHEP}\ }\textbf {\bibinfo {volume} {1610}},\
  \bibinfo {pages} {058} (\bibinfo {year} {2016}{\natexlab{b}})},\ \Eprint
  {http://arxiv.org/abs/1606.06312} {arXiv:1606.06312} \BibitemShut {NoStop}%
\bibitem [{\citenamefont {Catterall}\ and\ \citenamefont
  {Schaich}(2017)}]{Catterall:2016dzf}%
  \BibitemOpen
  \bibfield  {author} {\bibinfo {author} {\bibfnamefont {S.}~\bibnamefont
  {Catterall}}\ and\ \bibinfo {author} {\bibfnamefont {D.}~\bibnamefont
  {Schaich}},\ }\href {\doibase 10.1103/PhysRevD.96.034506} {\bibfield
  {journal} {\bibinfo  {journal} {Phys. Rev.}\ }\textbf {\bibinfo {volume}
  {D96}},\ \bibinfo {pages} {034506} (\bibinfo {year} {2017})},\ \Eprint
  {http://arxiv.org/abs/1609.08541} {arXiv:1609.08541 [hep-lat]} \BibitemShut
  {NoStop}%
\bibitem [{\citenamefont {He}\ \emph {et~al.}(2016)\citenamefont {He},
  \citenamefont {Wu}, \citenamefont {You}, \citenamefont {Xu}, \citenamefont
  {Meng},\ and\ \citenamefont {Lu}}]{He:2016sbs}%
  \BibitemOpen
  \bibfield  {author} {\bibinfo {author} {\bibfnamefont {Y.-Y.}\ \bibnamefont
  {He}}, \bibinfo {author} {\bibfnamefont {H.-Q.}\ \bibnamefont {Wu}}, \bibinfo
  {author} {\bibfnamefont {Y.-Z.}\ \bibnamefont {You}}, \bibinfo {author}
  {\bibfnamefont {C.}~\bibnamefont {Xu}}, \bibinfo {author} {\bibfnamefont
  {Z.~Y.}\ \bibnamefont {Meng}}, \ and\ \bibinfo {author} {\bibfnamefont
  {Z.-Y.}\ \bibnamefont {Lu}},\ }\href {\doibase 10.1103/PhysRevB.94.241111}
  {\bibfield  {journal} {\bibinfo  {journal} {Phys. Rev.}\ }\textbf {\bibinfo
  {volume} {B94}},\ \bibinfo {pages} {241111} (\bibinfo {year} {2016})},\
  \Eprint {http://arxiv.org/abs/1603.08376} {arXiv:1603.08376} \BibitemShut
  {NoStop}%
\bibitem [{\citenamefont {Schaich}\ and\ \citenamefont
  {Catterall}(2017)}]{Schaich:2017czc}%
  \BibitemOpen
  \bibfield  {author} {\bibinfo {author} {\bibfnamefont {D.}~\bibnamefont
  {Schaich}}\ and\ \bibinfo {author} {\bibfnamefont {S.}~\bibnamefont
  {Catterall}},\ }in\ \href
  {https://inspirehep.net/record/1631997/files/arXiv:1710.08137.pdf} {\emph
  {\bibinfo {booktitle} {{35th International Symposium on Lattice Field Theory
  (Lattice 2017) Granada, Spain, June 18-24, 2017}}}}\ (\bibinfo {year}
  {2017})\ \Eprint {http://arxiv.org/abs/1710.08137} {arXiv:1710.08137
  [hep-lat]} \BibitemShut {NoStop}%
\bibitem [{\citenamefont {You}\ \emph {et~al.}(2017)\citenamefont {You},
  \citenamefont {He}, \citenamefont {Xu},\ and\ \citenamefont
  {Vishwanath}}]{You:2017ltx}%
  \BibitemOpen
  \bibfield  {author} {\bibinfo {author} {\bibfnamefont {Y.-Z.}\ \bibnamefont
  {You}}, \bibinfo {author} {\bibfnamefont {Y.-C.}\ \bibnamefont {He}},
  \bibinfo {author} {\bibfnamefont {C.}~\bibnamefont {Xu}}, \ and\ \bibinfo
  {author} {\bibfnamefont {A.}~\bibnamefont {Vishwanath}},\ }\href@noop {} {\
  (\bibinfo {year} {2017})},\ \Eprint {http://arxiv.org/abs/1705.09313}
  {arXiv:1705.09313 [cond-mat.str-el]} \BibitemShut {NoStop}%
\bibitem [{\citenamefont {Banks}\ \emph {et~al.}(1982)\citenamefont {Banks},
  \citenamefont {Dothan},\ and\ \citenamefont {Horn}}]{Banks:1982iq}%
  \BibitemOpen
  \bibfield  {author} {\bibinfo {author} {\bibfnamefont {T.}~\bibnamefont
  {Banks}}, \bibinfo {author} {\bibfnamefont {Y.}~\bibnamefont {Dothan}}, \
  and\ \bibinfo {author} {\bibfnamefont {D.}~\bibnamefont {Horn}},\ }\href
  {\doibase 10.1016/0370-2693(82)90571-8} {\bibfield  {journal} {\bibinfo
  {journal} {Phys. Lett.}\ }\textbf {\bibinfo {volume} {B117}},\ \bibinfo
  {pages} {413} (\bibinfo {year} {1982})}\BibitemShut {NoStop}%
\bibitem [{\citenamefont {Bock}\ \emph {et~al.}(1992)\citenamefont {Bock},
  \citenamefont {Smit},\ and\ \citenamefont {Vink}}]{Bock:1992yr}%
  \BibitemOpen
  \bibfield  {author} {\bibinfo {author} {\bibfnamefont {W.}~\bibnamefont
  {Bock}}, \bibinfo {author} {\bibfnamefont {J.}~\bibnamefont {Smit}}, \ and\
  \bibinfo {author} {\bibfnamefont {J.~C.}\ \bibnamefont {Vink}},\ }\href
  {\doibase 10.1016/0370-2693(92)91049-F} {\bibfield  {journal} {\bibinfo
  {journal} {Phys. Lett.}\ }\textbf {\bibinfo {volume} {B291}},\ \bibinfo
  {pages} {297} (\bibinfo {year} {1992})},\ \Eprint
  {http://arxiv.org/abs/hep-lat/9206008} {arXiv:hep-lat/9206008 [hep-lat]}
  \BibitemShut {NoStop}%
\bibitem [{\citenamefont {Golterman}\ and\ \citenamefont
  {Smit}(1984)}]{Golterman:1984cy}%
  \BibitemOpen
  \bibfield  {author} {\bibinfo {author} {\bibfnamefont {M.~F.~L.}\
  \bibnamefont {Golterman}}\ and\ \bibinfo {author} {\bibfnamefont
  {J.}~\bibnamefont {Smit}},\ }\href {\doibase 10.1016/0550-3213(84)90424-3}
  {\bibfield  {journal} {\bibinfo  {journal} {Nucl. Phys.}\ }\textbf {\bibinfo
  {volume} {B245}},\ \bibinfo {pages} {61} (\bibinfo {year}
  {1984})}\BibitemShut {NoStop}%
\bibitem [{\citenamefont {van~den Doel}\ and\ \citenamefont
  {Smit}(1983)}]{vandenDoel:1983mf}%
  \BibitemOpen
  \bibfield  {author} {\bibinfo {author} {\bibfnamefont {C.}~\bibnamefont
  {van~den Doel}}\ and\ \bibinfo {author} {\bibfnamefont {J.}~\bibnamefont
  {Smit}},\ }\href {\doibase 10.1016/0550-3213(83)90401-7} {\bibfield
  {journal} {\bibinfo  {journal} {Nucl. Phys.}\ }\textbf {\bibinfo {volume}
  {B228}},\ \bibinfo {pages} {122} (\bibinfo {year} {1983})}\BibitemShut
  {NoStop}%
\bibitem [{\citenamefont {van Baal}\ and\ \citenamefont
  {Wipf}(2001)}]{vanBaal:2001jm}%
  \BibitemOpen
  \bibfield  {author} {\bibinfo {author} {\bibfnamefont {P.}~\bibnamefont {van
  Baal}}\ and\ \bibinfo {author} {\bibfnamefont {A.}~\bibnamefont {Wipf}},\
  }\href {\doibase 10.1016/S0370-2693(01)00856-5} {\bibfield  {journal}
  {\bibinfo  {journal} {Phys. Lett.}\ }\textbf {\bibinfo {volume} {B515}},\
  \bibinfo {pages} {181} (\bibinfo {year} {2001})},\ \Eprint
  {http://arxiv.org/abs/hep-th/0105141} {arXiv:hep-th/0105141 [hep-th]}
  \BibitemShut {NoStop}%
\bibitem [{\citenamefont {He}\ and\ \citenamefont {Guo}(2014)}]{He:2014eoa}%
  \BibitemOpen
  \bibfield  {author} {\bibinfo {author} {\bibfnamefont {Y.}~\bibnamefont
  {He}}\ and\ \bibinfo {author} {\bibfnamefont {H.}~\bibnamefont {Guo}},\
  }\href {\doibase 10.1016/j.physletb.2014.10.039} {\bibfield  {journal}
  {\bibinfo  {journal} {Phys. Lett.}\ }\textbf {\bibinfo {volume} {B739}},\
  \bibinfo {pages} {83} (\bibinfo {year} {2014})},\ \Eprint
  {http://arxiv.org/abs/1405.4089} {arXiv:1405.4089 [math-ph]} \BibitemShut
  {NoStop}%
\bibitem [{\citenamefont {BenTov}\ and\ \citenamefont
  {Zee}(2016)}]{BenTov:2015gra}%
  \BibitemOpen
  \bibfield  {author} {\bibinfo {author} {\bibfnamefont {Y.}~\bibnamefont
  {BenTov}}\ and\ \bibinfo {author} {\bibfnamefont {A.}~\bibnamefont {Zee}},\
  }\href {\doibase 10.1103/PhysRevD.93.065036} {\bibfield  {journal} {\bibinfo
  {journal} {Phys. Rev.}\ }\textbf {\bibinfo {volume} {D93}},\ \bibinfo {pages}
  {065036} (\bibinfo {year} {2016})},\ \Eprint
  {http://arxiv.org/abs/1505.04312} {arXiv:1505.04312 [hep-th]} \BibitemShut
  {NoStop}%
\bibitem [{\citenamefont {Fidkowski}\ and\ \citenamefont
  {Kitaev}(2010)}]{Fidkowski:2009dba}%
  \BibitemOpen
  \bibfield  {author} {\bibinfo {author} {\bibfnamefont {L.}~\bibnamefont
  {Fidkowski}}\ and\ \bibinfo {author} {\bibfnamefont {A.}~\bibnamefont
  {Kitaev}},\ }\href {\doibase 10.1103/PhysRevB.81.134509} {\bibfield
  {journal} {\bibinfo  {journal} {Phys. Rev.}\ }\textbf {\bibinfo {volume}
  {B81}},\ \bibinfo {pages} {134509} (\bibinfo {year} {2010})},\ \Eprint
  {http://arxiv.org/abs/0904.2197} {arXiv:0904.2197} \BibitemShut {NoStop}%
\bibitem [{\citenamefont {Eichten}\ and\ \citenamefont
  {Preskill}(1986)}]{Eichten:1985ft}%
  \BibitemOpen
  \bibfield  {author} {\bibinfo {author} {\bibfnamefont {E.}~\bibnamefont
  {Eichten}}\ and\ \bibinfo {author} {\bibfnamefont {J.}~\bibnamefont
  {Preskill}},\ }\href {\doibase 10.1016/0550-3213(86)90207-5} {\bibfield
  {journal} {\bibinfo  {journal} {Nucl. Phys.}\ }\textbf {\bibinfo {volume}
  {B268}},\ \bibinfo {pages} {179} (\bibinfo {year} {1986})}\BibitemShut
  {NoStop}%
\bibitem [{\citenamefont {You}\ and\ \citenamefont {Xu}(2015)}]{You:2014vea}%
  \BibitemOpen
  \bibfield  {author} {\bibinfo {author} {\bibfnamefont {Y.-Z.}\ \bibnamefont
  {You}}\ and\ \bibinfo {author} {\bibfnamefont {C.}~\bibnamefont {Xu}},\
  }\href {\doibase 10.1103/PhysRevB.91.125147} {\bibfield  {journal} {\bibinfo
  {journal} {Phys. Rev.}\ }\textbf {\bibinfo {volume} {B91}},\ \bibinfo {pages}
  {125147} (\bibinfo {year} {2015})},\ \Eprint {http://arxiv.org/abs/1412.4784}
  {arXiv:1412.4784} \BibitemShut {NoStop}%
\end{thebibliography}%

\end{document}